\documentclass[toc]{PoS}
\usepackage{cite,amsmath}

\newcommand{\pisiSE}{$\Pi\Sigma^*$}

\newcommand{\KK}{\mathbb{K}}
\newcommand{\FF}{\mathbb{F}}
\newcommand{\NN}{\mathbb{N}}
\newcommand{\GG}{\mathbb{G}}

\newcommand{\dfield}[2]{({#1},{#2})}
\newcommand{\const}[2]{{\rm const}_{#2}{#1}}

\def\z#1{{\zeta_{#1}}}
\newcommand{\SigmaP}{{\sf Sigma}}
\newcommand{\Summer}{{\sf Summer}}

\newcommand{\Maple}{{\sf Maple}}
\newcommand{\Mathematica}{{\sf Mathematica}}
\newcommand{\Form}{{\sf Form}}

\title{Feynman integrals and difference equations}

\ShortTitle{Feynman integrals and difference equations}

\author{\speaker{S. Moch}\\
        Deutsches Elektronen--Synchrotron, DESY   \\
        Platanenallee 6, D--15738 Zeuthen, Germany\\
        E-mail: \email{sven-olaf.moch@desy.de}
}

\author{C. Schneider\\
        Research Institute for Symbolic Computation,
        Johannes Kepler University Linz, \\
        Altenberger Str. 69, A-4040, Austria\\
        E-mail: \email{carsten.schneider@risc.uni-linz.ac.at}
}

\abstract{
We report on the calculation of multi-loop Feynman integrals for single-scale problems
by means of difference equations in Mellin space.
The solution to these difference equations in terms of harmonic sums can be
constructed algorithmically over difference fields, the so-called \pisiSE-fields.
We test the implementaion of the \Mathematica\ package \SigmaP\
on examples from recent higher order perturbative calculations in Quantum Chromodynamics.
}

\FullConference{XI International Workshop on Advanced Computing and Analysis Techniques in Physics Research\\
         April 23-27 2007, \, Amsterdam, the Netherlands}

\begin{document}

\section{Introduction}

In quantum field theory the perturbative calculation of a given scattering amplitude or cross section
requires the evaluation of Feynman diagrams.
Especially at higher orders this is a difficult task and over the last years,
a large variety of methods has been devised to deal with the problem of calculating Feynman
integrals, see e.g.~\cite{Smirnov:2004ym} and references therein.
The necessary integration over the propagators of the virtual or unobserved particles
are typically carried out in momentum space and divergent integrals are regularized dimensionally
by shifting from 4 to $D = 4-2\epsilon$ dimensions of space-time.
For a given Feynman integral the main task is then the derivation of an analytical expression
in terms of known functions with well-defined properties,
which at the same time permits a Laurent expansion in the small parameter $\epsilon=(D-4)/2$.

In these proceedings we want to focus on particular progress in this direction
through the systematic and efficient approach to solve difference equations.
To that end, we  start by briefly stating the physics case and present
the necessary mathematical definitions. Then we provide explicit examples from
recent higher order perturbative calculations in Quantum Chromodynamics (QCD)
and finish with a summary and an outlook.

\subsection{Setting the stage}

For a given scattering process the Feynman integrals are classified by the
number $n$ of external legs ($n$-point functions), by the number of independent loops $l$
and by the topology of the associated graph.
Initially, the Feynman integrals will appear as tensor integrals $I^{\mu_1, \mu_2,\dots}$
where $\mu_i$ denote Lorentz indices.
Subsequently tensor integrals can be mapped to scalar integrals by numerous
methods, thus we are dealing with expressions
  \begin{eqnarray}
    \label{eq:Feynman-int-def}
    I(D;\nu_1, \dots, \nu_n) \,=\, \int\, d^Dp_1\,\dots\, d^Dp_l\,
    {1 \over (p_1^2-m_1^2)^{\nu_1} \, \dots\,  (p_n^2-m_n^2)^{\nu_n}}\, ,
  \end{eqnarray}
where the momenta $p_i$, $i=l+1,\dots,n$, are related by energy-momentum conservation $p_i = f(p_1,\dots,p_l)$,
and $m_i$ denote the masses of the associated particles.
The powers of the propagators are $\nu_i$ and $D$ is the (complex) space-time dimension.
In the most general case
Feynman integrals such as in Eq.~\eqref{eq:Feynman-int-def} may depend on multiple scales,
for instance the masses $m_i$ but also the non-vanishing scalar products of external momenta.

In the following, we will focus on perturbative calculations to higher orders
in massless QCD for single-particle inclusive observables.
These depend on a single scaling variable $x$ only, with $x \in [0,1]$.
Prominent examples are structure functions in deep-inelastic scattering (DIS),
fragmentation functions in electron-positron ($e^+e^-$) annihilation
or the total cross sections for the Drell-Yan process or Higgs production at hadron colliders.
All these quantities can be solved directly and systematically in Mellin $N$-space,
an approach that has been used in the past within the framework of the operator product expansion
(and exploiting the optical theorem) in DIS~\cite{Kazakov:1987jk,Moch:1999eb}.
More recently, also innovative extensions to other kinematics have been considered~\cite{Mitov:2005ps}.
For a generic observable ${\cal O}(x)$, we can write
\begin{eqnarray}
\label{eq:Mellin-N}
{\cal O}(N)
  &\!=\!&
  \int_0^1 dx\, x^{N-1} {\cal O}(x)
\nonumber\\
  &\!=\!&
  \int_0^1 dx\, x^{N-1} \int {\rm dPS}^{(m)}
    \bigl|M(\{\mbox{in}\} \to \{\mbox{out}\}) \bigr|^2 \delta(x-f)
\nonumber\\
  &\!=\!&
  \int {\rm dPS}^{(m)}
    \bigl|M(\{\mbox{in}\} \to \{\mbox{out}\}) \bigr|^2 {1 \over f^{-N+1}}
\, ,
\end{eqnarray}
where the first line defines the Mellin transform.
Subsequently, we express ${\cal O}(x)$ through the square of the
scattering amplitude $M(\{\mbox{in}\} \to \{\mbox{out}\})$
for a given set of incoming and outgoing particles and ${\rm dPS}^{(m)}$
denotes the integration measure of the Lorentz invariant phase space.
The invariant $f$ depends on internal and external momenta and introduces
the Mellin-$N$ dependence.
As an upshot, the observables are mapped to a discrete set of variables (positive integer Mellin $N$).

Once the steps in Eq.~\eqref{eq:Mellin-N} are accomplished, the scalar integrals can be
reduced algebraically to so-called master integrals.
The reduction algorithms are based on integration-by-parts, see e.g.~\cite{Smirnov:2004ym},
\begin{eqnarray}
  \label{eq:ibp-def}
  0 \,=\, \int\, d^Dp_1\,\dots\, d^Dp_l\,
  \left( {p_i}_{\mu} {\partial \over \partial p_j^{\mu}} \right) \,
{1 \over (p_1^2)^{\nu_1} \, \dots\,  (p_n^2)^{\nu_n}}\, {1 \over f^{-N+1}}\, ,
\end{eqnarray}
where $p_i$ and $p_j$ denote any of the loop momenta and the Mellin-$N$
dependence is implicit through the invariant $f$, cf. Eq.~\eqref{eq:Mellin-N}.
Upon resolving the constraints from $f$ the reductions in Mellin space act
on monomials like $(p_i^2)^{-\nu_i + N}$ and give rise to systems of linear equations
which can be solved in terms of a (small) set of master integrals.
This step is automatized by using suitable (customized) computer algebra programs
although in practice limitations arise at higher orders through the
excessively large size of the systems of linear equations which need to be considered.

Due to the explicit dependence on the Mellin variable $N$
the master integrals themselves are functions of $N$.
Their functional dependence on $N$ is completely determined by the difference equations
they satisfy.
These difference equations in $N$ are obtained as well from the solutions
to the integration-by-parts reductions and they are the central topic
of the present proceedings.

With the help of Eq.~\eqref{eq:Mellin-N} and the solutions to the
integration-by-parts reductions, we are thus in a position to express a given Feynman
integral $I(N)$ in Mellin space in a recursive manner,
\begin{eqnarray}
\label{eq:k-step-diff}
  a_0(N) I(N) + a_1(N) I(N-1) + \dots + a_k(N) I(N-k) \,=\, X(N)
\, ,
\end{eqnarray}
which defines a difference equation of order $k$
with the parametric dependence on $\epsilon$ being implicit in Eq.~\eqref{eq:k-step-diff}.
The functions $a_i(N)$ are polynomials in $N$ (and $\epsilon$), sometimes they factorize 
linearly in terms of the type $N + m + n \epsilon$
with integer $m, n$.
$X(N)$ denotes the inhomogeneous part which collects the simpler integrals and
in a Laurent series in $\epsilon$ it is typically composed of
harmonic sums $S_{\vec{m}}$ (${\vec{m}}=m_1,\dots,m_k$) of weight $w=|m_1|+\dots+|m_k|$
and, possibly, in combination with values of the $\zeta$-function and powers of $N$.

From a practical view-point in quantum field theory, an approach like
Eq.~\eqref{eq:k-step-diff} allows for easy checking at fixed values of the Mellin moment $N$.
The analytical solution to Eq.~\eqref{eq:k-step-diff} requires concepts and algorithms
of symbolic summation (discussed e.g. in~\cite{Moch:2005uq}) and has been
possible in all cases we have encountered in QCD calculations~\cite{Moch:2004pa,Vogt:2004mw,Vermaseren:2005qc,Mitov:2006wy}.
However, the recurrence relations can also be embedded in a mathematical framework,
to which we will turn in the following.

\section{Solving recurrence relations in difference fields}

The Mellin space approach to Feynman integrals at higher orders which we have
briefly sketched above leads us to the following key problem:

\textbf{Given}
sequences $a_0(k),\dots,a_m(k)$ and $f_1(k),\dots,f_d(k)$, \textbf{find} all constants $c_1,\dots,c_d$, free of $k$, and all $g(k)$
such that the parameterized linear difference equation
\begin{equation}
\label{Equ:PLDE}
a_m(k)g(k+m)+a_{m-1}(k)g(k+m-1)+\dots +a_0(k)g(k) = c_1f_1(k)+\dots+c_df_d(k)
\end{equation}
holds.

Note that this problem covers various prominent subproblems.
Specializing to $d=1$, this is nothing else than recurrence
solving, i.e. we are back to Eq.~\eqref{eq:k-step-diff}.
Taking $m=1$ with $a_0=-1$ and $a_1=1$ we get parameterized
telescoping; in particular, given a bivariate sequence $f(n,k)$, one
can set $f_i(k):=f(n+i-1,k)$ which corresponds in the hypergeometric
case to Zeilberger's creative telescoping~\cite{Zeilberger:91}. Finally, choosing $d=m=1$
with $a_0=-1$, $a_1=1$, we arrive at telescoping; for more details and applications of these summation principles see~\cite{Bierenbaum:2007zu}.

The algorithmic solution of Eqs.~\eqref{eq:k-step-diff} or \eqref{Equ:PLDE}
requires algorithms for symbolic summation which are typically implemented in computer
algebra systems like \Maple, \Mathematica\ or \Form~\cite{Vermaseren:2000nd},
the latter being most advantageous if large expressions are involved.
Specific implementations that deal with symbolic summation are e.g. \Summer~\cite{Vermaseren:1998uu}
or the recent summation package \SigmaP~\cite{Schneider:07a}, which can handle
problem~\eqref{Equ:PLDE}, and therefore telescoping, creative
telescoping and recurrence solving, if the $a_0,\dots,a_m$ and $f_1,\dots,f_d$ are given
by expressions in terms of indefinite nested sums and products.
More precisely, \SigmaP\ translates such sum--product expressions into
difference fields, the so-called \pisiSE-fields, and solves the
given problem~\eqref{Equ:PLDE} there.

\subsection{The construction of difference fields}

Given an equation of the form~\eqref{Equ:PLDE}, one can represent the coefficients $a_i$ and
the inhomogeneous parts $f_i$ in the so-called \pisiSE-fields~\cite{Karr:81}.

\medskip

\noindent{\it Definition.} Let $\FF$ be a field with characteristic
$0$ and let $\sigma$ be a field automorphism of $\FF$. Then
$(\FF,\sigma)$ is called a difference field; the constant field of
$\FF$ is defined by $\const{\FF}{\sigma}=\{f\in\FF|\sigma(f)=f\}$. A difference
field $(\FF,\sigma)$ with constant field $\KK$ is called a
\pisiSE-field if $\FF=\KK(t_1,\dots,t_e)$ where for all $1\leq i\leq
e$ each $\FF_i=\KK(t_1,\dots,t_i)$ is a transcendental field
extension of\footnote{We set $\FF_0=\KK$.} $\FF_{i-1}=\KK(t_1,\dots,t_{i-1})$
and $\sigma$ has the property that $\sigma(t_i)=a\,t_i$ or
$\sigma(t_i)=t_i+a$ for some $a\in\FF_{i-1}$.

\medskip

\noindent\textit{Remark.} In order to transform, e.g., nested sums in such \pisiSE-fields, one exploits the following important fact. Suppose we are given a sum
$T(n)=\sum_{k=1}^nF(k-1)$ where we have represented $F(n)$ in a
\pisiSE-field $\dfield{\FF}{\sigma}$ with $f\in\FF$. Then, one can
either express $T(n)$ in $\FF$ by solving the telescoping
problem: \textbf{ Find} $g\in\FF$ with
\begin{equation}\label{Equ:DFTele}
\sigma(g)-g=f.
\end{equation}
Namely, if we find such a $g$, we can rephrase $T(n)$ by
$t:=g+c$ for some $c\in\const{\FF}{\sigma}$, in particular, the
shift behavior $S(n+1)=S(n)+f(n)$ is reflected by $\sigma(t)=t+f$.\\
Otherwise, if we fail to find such a $g$, we adjoin the sum in form  of the transcendental field extension $\FF(t)$ where the field automorphism is extended from $\FF$ to $\FF(t)$ by the shift behavior $\sigma(t)=t+f$. Then by Karr's remarkable result~\cite{Karr:81} it follows that the constant field is not enlarged, i.e., $\const{\FF(t)}{\sigma}=\const{\FF}{\sigma}$. In other words, $\dfield{\FF(t)}{\sigma}$ is again a \pisiSE-field.

\medskip

In a nutshell, one either can represent a given sum in the already constructed field $\FF$ by solving the telescoping problem~\eqref{Equ:DFTele}, or otherwise, one can adjoin the sum in form of a transcendental extension; for products we refer to~\cite{Schneider:05c}.
Finally, we emphasize that this construction process is completely
algorithmically, see Sec.~\ref{Sec:PLDEDF} and therefore the translation
mechanism can be carried out automatically, e.g. in the \Mathematica\ package \SigmaP.

\subsection{Solving linear difference equations in the ground field}\label{Sec:PLDEDF}

Suppose we are given Eq.~\eqref{Equ:PLDE} and we have rephrased the
$a_0,\dots,a_m$ and $f_1,\dots,f_d$ in a \pisiSE-field
$\FF=\KK(t_1,\dots,t_e)$; in short we call such a difference field also the
coefficient field of Eq.~\eqref{Equ:PLDE}. Then problem~\eqref{Equ:PLDE} can be rephrased as follows:\\
\textbf{Find} all $c_1,\dots,c_d\in\KK$ and $g\in\FF$ such that
\begin{equation}\label{Equ:PLDEDF}
a_m\sigma(g)+a_{m-1}\sigma^{m-1}(g)+\dots+a_0g=c_1f_1+\dots+c_df_d.
\end{equation}

\medskip

\noindent{\it Remark.} Eq.~\eqref{Equ:PLDEDF} can be solved in $\FF$ by solving several such problems in the subfield~$\GG:=\KK(t_1,\dots,t_{e-1})$. Namely, we arrive at the following reduction~\cite{Schneider:05a}; we set $t:=t_e$, i.e., $\FF=\GG(t)$:

\noindent{\bf Reduction I} (denominator bounding). Compute a nonzero polynomial
$d\in\GG[t]$ such that for all $c_i\in\KK$ and $g\in\GG(t)$
with Eq.~\eqref{Equ:PLDEDF} we have $dg\in\GG[t]$. Then it follows
that
\begin{equation}\label{Equ:ParaTeleDen}
\frac{a_0}{d}g'+\dots+\frac{a_m}{\sigma^{m}(d)}\sigma^{m}(g')=c_1f_1+\dots+c_df_d
\end{equation}
for $g'\in\GG[t]$ if and only if Eq.~\eqref{Equ:PLDEDF} with $g=g'/d$.

\noindent{\bf Reduction II} (degree bounding).
Given such a denominator bound, it suffices to look only for
$c_i\in\KK$ and polynomial solutions $g\in\GG[t]$
with Eq.~\eqref{Equ:PLDEDF}.
Next, we compute a degree bound $b\in\NN_0$ for these polynomial solutions

\noindent{\bf Reduction III} (polynomial degree reduction). Given such a
degree bound one looks for $c_i\in\KK$ and $g_i\in\GG$ such
that Eq.~\eqref{Equ:PLDEDF} holds for $g=\sum_{i=0}^b g_it^i$.
This can be achieved as follows. First derive the
possible leading coefficients $g_b$ by solving a specific parameterized linear difference equation given
in $\GG$, then plug its solutions into Eq.~\eqref{Equ:PLDEDF}
and recursively look for the remaining solutions $g=\sum_{i=0}^{b-1} g_it^i$.
Thus one can derive the solutions of Eq.~\eqref{Equ:PLDEDF} given in $\GG(t)$ by
solving several such parameterized linear difference equations given in the ground field $\GG$.

\medskip

Together with results from~\cite{Karr:81,Bron:00,Schneider:04b,Schneider:05b} it has been shown in~\cite{Schneider:05a}
that this reduction leads to a complete algorithm for problem~\eqref{Equ:PLDEDF} with $m=1$; as result we obtain a streamlined and simplified version of Karr's algorithm~\cite{Karr:81} whose main interest is to solve the telescoping problem~\eqref{Equ:DFTele}. 
Moreover, it is shown that this reduction also delivers a method that eventually produces
all solutions for recurrences of higher order $m\geq2$.

\subsection{d'Alembertian solution}

Now suppose that we have rephrased Eq.~\eqref{Equ:PLDE} in the form~\eqref{Equ:PLDEDF} for a particular coefficient field $\FF$. Then
$\FF$ is usually too small to find all the required solutions.
Therefore, the following more general approach is helpful:\\
\textbf{Find} all  solutions of the form
\begin{equation}\label{DefOfDAlembertSol}
h(n)\sum_{k_1=0}^nb_1(k_1)\sum_{k_2=0}^{k_2}
b_2(k_2)\dots\sum_{k_s=0}^{k_{s-1}}b_s(k_s)
\end{equation}
where the $b_i(k_i)$ and $h(n)$ are represented in $\FF$ or they are defined as products over elements from $\FF$. Such type of solutions are called
d'Alembertian solutions~\cite{Abramov:94}, which are a subclass of Liouvillian solutions~\cite{Singer:99}.

\medskip

\noindent\textit{Remark.} The d'Alembertian solutions are obtained by
factorizing the linear difference equation as much as possible into first
order linear right factors over the given difference field/ring.  Then each
factor corresponds basically to one indefinite summation quantifier;
see~\cite{Abramov:94,Schneider:T01}. We remark that finding such a linear
right hand factor is equivalent to finding a product solution of the
recurrence in its coefficient domain. For the rational case, i.e.,
$a_i(k),f_i(k)\in\KK(k)$, this problem has been solved in~\cite{Petkov:92}. A
general version for the \pisiSE-field situation, which uses the methods
described in Sec.~\ref{Sec:PLDEDF} as subroutines, has been developed in
the summation package~\SigmaP; see also~\cite{Schneider:T01,Schneider:07a}.

\medskip

We stress the following important aspect: one usually needs simplified representations 
of the solutions~\eqref{DefOfDAlembertSol} for further treatment.
If the solutions are given in form of harmonic sums, one can derive compact
representations by using the algebraic relations of harmonic sums (see e.g.~\cite{Blumlein:2003gb}), 
or alternatively, use \SigmaP\ for this task.  
Ongoing research (see e.g.~\cite{Schneider:05f}) is dedicated to the
computation of sum representations of the type~\eqref{DefOfDAlembertSol} with optimal nested depth.

\medskip

It is worth pointing out, that within the formalism of \pisiSE-fields,
a number of extensions/generalizations can be treated (e.g. in \SigmaP).
These include algebraic objects like $(-1)^n$.
Note that such elements cannot be represented in fields,
but only in rings with zero-divisors, like $(1+(-1)^n)(1-(-1)^n)=0$;
for more details see~\cite{Schneider:T01}.
One can also consider radical objects like $\sqrt{n}$ or
free/generic sequences $X_n$, for which
the corresponding algorithms are presented in~\cite{Schneider:07f}
and in~\cite{Schneider:06d,Schneider:06e}, respectively.

Summarizing, given an equation of the form~\eqref{Equ:PLDE}, the mathematical
framework of \pisiSE-fields puts us in a position to solve the
corresponding recurrence relations in terms of indefinite nested sums
and products over the given coefficient field $\FF$.
This class covers harmonic sums $S_{\vec{m}}$ and is therefore
sufficient for solutions of difference equations for
single scale Feynman integrals in the Mellin space approach.

\section{Examples of Feynman integrals}

\begin{figure}
  \centering
  \includegraphics[width=8.0cm,angle=0]{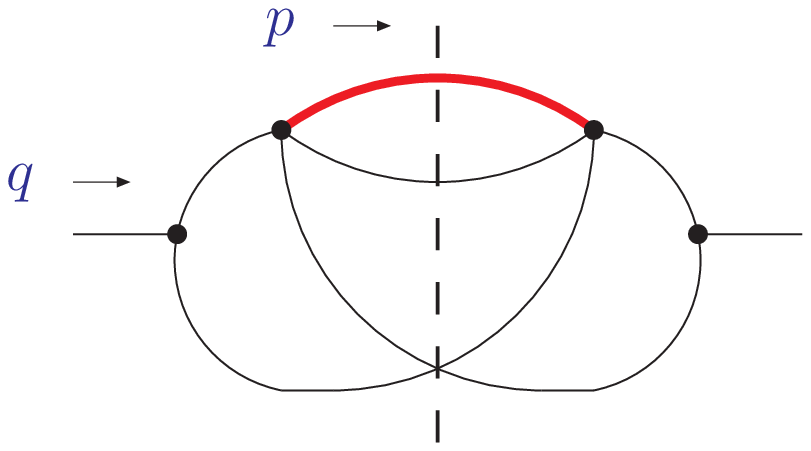}
  \includegraphics[width=7.0cm,angle=0]{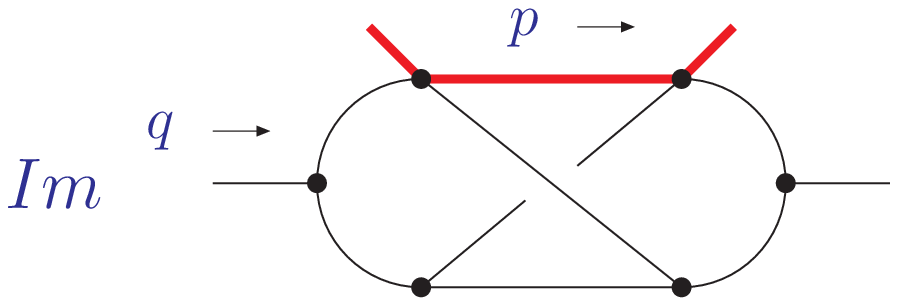}
  \caption{
  \label{fig:diags}
  Examples of Feynman integrals from $e^+e^-$-annihilation (left) and
  deep-inelastic scattering (right). All propagators have unit power.}
\end{figure}

Here we present two examples from QCD calculations to two- and three loops
for DIS structure functions for single hadron inclusive
$e^+e^-$-annihilation~\cite{Moch:2004pa,Vogt:2004mw,Vermaseren:2005qc,Mitov:2006wy}.
The relevant diagrams are displayed in Fig.~\ref{fig:diags}
and the respective difference equations for the Feynman integrals $I(N)$
are of second and third order.
They could be solved by matching to an ansatz of the type
\begin{eqnarray}
\label{eq:ansatz}
I(N) & = &
   \sum\, a(j,{\vec{m}})\, \epsilon^j\, S_{\vec{m}}(N)
 + \sum\, b(j,{\vec{m}},k)\, \epsilon^j\, \zeta_k\, S_{\vec{m}}(N) \, ,
\end{eqnarray}
were the unknown coefficients $a$ and $b$ were then determined
with the \Summer\ package~\cite{Vermaseren:1998uu} by inserting
Eq.~\eqref{eq:ansatz} in the recursion relation~\eqref{eq:k-step-diff}.
This approach, of course, rests entirely on the fact
that the solution for the Feynman integrals $I(N)$
as a Laurent series in $\epsilon$ is within the space
spanned by harmonic sums $\displaystyle S_{\vec{m}}(N)$ and
Riemann $\zeta$-values $\zeta_i$.
For all Feynman integrals $I(N)$ which were determined by higher order
difference equations we found this condition to be fulfilled.

An alternative path to the solution, based on the mathematical
framework developed in Sec.~\ref{Sec:PLDEDF} above, is provided by the \SigmaP\ package.
This we want to discuss next.

\subsection{$e^+e^-$-annihilation}

The example which stems from single hadron inclusive $e^+e^-$-annihilation
is a phase space integral for a decay process $1 \to 4$
in one-particle inclusive kinematics, see Fig.~\ref{fig:diags} on the left.
The scalar diagram describes the decay of a particle with initial momentum $q$
according to $q \to p_1 + p_2 + p_3 + p$
and the dashed vertical line denotes the final-state cut for the production of particles.
Our example is the master integral ${\mbox{R}}_2(N)$ from the
so-called {\it real-real}-emission and it depends
on the dimensionless variable $x_E$ which is the scaled momentum fraction
\begin{equation}
  \label{eq:xdefT}
x_E = {2 p \cdot q \over q \cdot q}\, , \quad\quad\quad  0 \le x_E \le 1\, .
\end{equation}

The solution of the integration-by-parts identities leads to a
difference equation of second order~\cite{Mitov:2006wy},
which reflects the underlying symmetries of the Feynman diagram.
We find
\begin{eqnarray}
\label{eq:rm2-n}
{\lefteqn{
         (N + 1 - 2 \* \epsilon) \* (N + 2 - 6 \* \epsilon)  \*  {\mbox{R}}_2(N)
       - (N-1) \* (N - 4 \* \epsilon)  \*  {\mbox{R}}_2(N-2)
\, = \,
}}
\nonumber\\
&& \qquad\qquad\qquad\qquad
         2 \* {{(1-\epsilon) \* (1 - 3 \* \epsilon) \* (2 - 3 \* \epsilon) \*
           (2 \* N + 1 - 6 \* \epsilon)} \over {(N - 3 \* \epsilon) \* (N + 1 - 3 \* \epsilon)}}\,
       \* {\mbox{R}}_1(N-2)
\: \: ,
\end{eqnarray}
where ${\mbox{R}}_1$ denotes the inhomogeneous part.
For a complete solution, we also have to provide the initial conditions ${\mbox{R}}_2(0)$ and ${\mbox{R}}_2(1)$.
The explicit expressions are rather lengthy, in particular the higher powers in the Laurent
series in $\epsilon$ and we refer the reader to~\cite{Mitov:2006wy} for details.

We have provided \SigmaP\ with the difference equation~\eqref{eq:rm2-n} and
the expression for ${\mbox{R}}_1$ in terms of harmonic sums.
Subsequently \SigmaP\ was able to solve for ${\mbox{R}}_2(N)$ and we have found
complete agreement with the published result after insertion of the
initial conditions ${\mbox{R}}_2(0)$ and ${\mbox{R}}_2(1)$.

\subsection{Deep-inelastic scattering}

During the computation of the three-loop QCD corrections to the DIS structure functions $F_2$ and $F_L$
we have encountered the Feynman integral shown in Fig.~\ref{fig:diags} (right).
The scalar diagram in the DIS case has external momenta $p$ and $q$
and is of the non-planar topology ${\mbox{NO}}_{22}$ with unit powers of the propagators.
Due to the framework of the operator product expansion and the optical theorem
applied in DIS, we are specifically interested in the imaginary part of ${\mbox{NO}}_{22}$.
In momentum space it depends on the scaling variable $x_B$,
\begin{equation}
  \label{eq:xdefS}
x_B = {- q \cdot q \over 2 p \cdot q }\, , \quad\quad\quad  0 \le x_B \le 1\, .
\end{equation}

In Mellin space the associated difference equation is of third order and
very difficult~\cite{Vermaseren:2005qc}.
It has been obtained by means of a systematic solution of the
integration-by-parts identities and is given by
\begin{eqnarray}
\label{eq:no22-n}
&&  -{\mbox{NO}}_{22}(N)\*N\*(N-2)\*(3\*N-5)\*(N^2-1)
\nonumber \\
&&  +{\mbox{NO}}_{22}(N-1)\*N\*(3\*N^4-20\*N^3+45\*N^2-40\*N+12)
\nonumber \\
&&  +{\mbox{NO}}_{22}(N-2)\*N\*(3\*N^4-17\*N^3+35\*N^2-31\*N+10)
\nonumber \\
&&  -{\mbox{NO}}_{22}(N-3)\*(N-1)\*(N-2)^2\*(3\*N-2)\*(N-3)
  \,=\,
2\*{\mbox{X}}(N)\, ,
\end{eqnarray}
where we have put $\epsilon = 0$, since the integral ${\mbox{NO}}_{22}$ does
not contain any divergences.
It is completely finite and to the accuracy needed in the three-loop
calculation~\cite{Vermaseren:2005qc} positive powers in $\epsilon$ were not necessary.

For the solution of Eq.~\eqref{eq:no22-n},
we have to provide the inhomogeneous part ${\mbox{X}}$ (which is also finite),
\begin{eqnarray}
\label{eq:Xinhom-n}
  {\mbox{X}}(N) &=& 
            32 \* \z3
          + 56 \* {S_{-3}(N-3) \over N-3}
          - 8 \* {S_{-3}(N-2) \over N-2}
          + 20 \* S_{-3}(N)
          + 108 \* {S_{-2}(N-3) \over N-3}
          + 56 \* {S_{-2}(N-3) \over (N-3)^2}
\nonumber\\
&& 
          - 36 \* {S_{-2}(N-2) \over N-2}
          - 8 \* {S_{-2}(N-2) \over (N-2)^2}
          + 12 \* S_{-2}(N)
          + 12 \* {S_{2}(N-2) \over N-2}
          + 16 \* {S_{2}(N-2) \over (N-2)^2}
\nonumber\\
&& 
          - 12 \* {S_{2}(N-1) \over N-1}
          + 8 \* {S_{2}(N-1) \over (N-1)^2}
          - 12 \* S_{2}(N)
          - 16 \* {S_{3}(N-2) \over N-2}
          - 8 \* {S_{3}(N-1) \over N-1}
          - 4 \* S_{3}(N)
\nonumber\\
&& 
          + 84 \* {\z3 \over N-3}
          + 24 \* {1 \over N-2}
          - 4 \* {\z3 \over N-2}
          + 92 \* {1 \over (N-2)^2}
          + 56 \* {1 \over (N-2)^3}
          - 48 \* {1 \over N-1}
          + 4 \* {\z3 \over N-1}
\nonumber\\
&& 
          - 100 \* {1 \over (N-1)^2}
          - 36 \* {1 \over (N-1)^3}
          + 48 \* {1 \over N}
          + 76 \* {1 \over N^2}
          + 20 \* {1 \over N^3}
       + (-1)^{N}  \*  \biggl(
          - 24 \* \z3
          - 72 \* N \* \z3
\nonumber\\
&& 
          - 56 \* {S_{-3}(N-3) \over N-3}
          - 24 \* {S_{-3}(N-2) \over N-2}
          + 8 \* {S_{-3}(N-1) \over N-1}
          - 16 \* S_{-3}(N)
          - 48 \* S_{-3}(N) \* N
\nonumber\\
&& 
          - 108 \* {S_{-2}(N-3) \over N-3}
          - 56 \* {S_{-2}(N-3) \over (N-3)^2}
          - 24 \* {S_{-2}(N-2) \over N-2}
          + 8 \* {S_{-2}(N-2) \over (N-2)^2}
          + 12 \* {S_{-2}(N-1) \over N-1}
\nonumber\\
&& 
          - 8 \* {S_{-2}(N-1) \over (N-1)^2}
          - 16 \* S_{-2}(N)
          - 48 \* S_{-2}(N) \* N
          - 84 \* {\z3 \over N-3}
          - 24 \* {1 \over N-2}
          - 36 \* {\z3 \over N-2}
\nonumber\\
&& 
          - 92 \* {1 \over (N-2)^2}
          - 56 \* {1 \over (N-2)^3}
          + 12 \* {\z3 \over N-1}
          + 12 \* {1 \over (N-1)^2}
          + 20 \* {1 \over (N-1)^3}
          - 12 \* {1 \over N^2}
          - 20 \* {1 \over N^3}
          \biggr)
\, ,
\nonumber\\
&& 
\end{eqnarray}
along with the initial conditions ${\mbox{NO}}_{22}(0)$, ${\mbox{NO}}_{22}(1)$
and ${\mbox{NO}}_{22}(2)$.
Here, $\zeta_i$ denotes the Riemann $\zeta$-function at value $i$.

We have again used \SigmaP\ for solving the difference
equation~\eqref{eq:no22-n}, given the expression for ${\mbox{X}}$ from Eq.~\eqref{eq:Xinhom-n}
and \SigmaP\ has provided us with the correct analytical answer for ${\mbox{NO}}_{22}$
once also the initial conditions ${\mbox{NO}}_{22}(0)$, ${\mbox{NO}}_{22}(1)$ and ${\mbox{NO}}_{22}(2)$
had been supplied.
The solution is given by the following very compact expression,
\begin{eqnarray}
\displaystyle
  \label{eq:no22-sol}  
{\mbox{NO}}_{22}(N)
&=&\frac{2 \left(6 \z3 +
  (-1)^N \left(6 \z3 -5 N^2
  \z5 \right)\right)}{N^2 (N+1)} \nonumber\\
&&-\frac{8 \left(1+(-1)^N\right)
  S_{-5}(N)}{N+1}-\frac{4 \left(1+(-1)^N\right)
  S_5(N)}{N+1}+S_2(N) \left(\frac{4 S_3(N)}{N+1}-\frac{4
  \z3 }{N+1}\right) \nonumber\\
&&+S_{-3}(N) \left(\frac{8 \left(1+(-1)^N\right)}{N^2
  (N+1)}-\frac{4 \left(2+(-1)^N\right) S_{-2}(N)}{N+1}-\frac{4
  S_2(N)}{N+1}\right) \nonumber\\
&&+S_{-2}(N) \left(\frac{-12 \z3  N^3+(-1)^N
  \left(8-8 N^3 \z3 \right)+8}{N^3 (N+1)}-\frac{\left(4+8
  (-1)^N\right) S_3(N)}{N+1}\right) \nonumber\\
&&+\frac{8 S_{-3,-2}(N)}{N+1}+
  \frac{\left(4-4(-1)^N\right)S_{-3,2}(N)}{N+1}+\frac{\left(4+12
(-1)^N\right) S_{-2,3}(N)}{N+1}-\frac{8S_{2,3}(N)}{N+1}
\, ,
\nonumber\\
&&
\end{eqnarray}
where we have eliminated a number of nested sums in favor of products of harmonic sums through
their product algebra, leaving only four independent sums of depth two.

\section{Summary}

In these proceedings, we have presented a short introduction to the
problem of calculating Feynman integrals at multiple loops.
For single-scale problems we have sketched how to formulate the problem in Mellin space,
how to obtain algebraic reductions for the loop integrals and, finally,
how these reductions lead to difference equations in Mellin space.
While the underlying physics problem usually places certain restrictions on
the difference equations, which allow for their solutions to
be expressed in terms of harmonic sums, and hence to be constructed with a corresponding ansatz,
it is advantageous to reconsider the problem in a more general mathematical framework.

We have reported on progress in this direction by exploiting algebraic properties
of difference fields, the so-called \pisiSE-fields.
The algorithmic construction of analytical solutions to the recursion relations
for Feynman integrals over difference fields as, for instance, implemented
in the \Mathematica\ package \SigmaP\ provides great calculational advantages.
To that end, we have tested the package \SigmaP\ on two examples of Feynman integrals
which had appeared in recent perturbative higher-order QCD computations
for $e^+e^-$-annihilation and deep-inelastic scattering.
With the improved mathematical machinery both examples have been re-evaluated
with the package \SigmaP\ and we have found agreement.

In the future more generalizations or extensions are conceivable.
From the need to consider physics problems and Feynman integrals depending on multiple scales
one arrives at generalized sums~\cite{Moch:2001zr},
which in turn are connected to multiple and harmonic polylogarithms~\cite{Goncharov,Remiddi:1999ew}.
A related problem is the expansion of (generalized) hypergeometric functions
around integer or rational values in a small parameter~
These problems lead to recursion relations similar to Eqs.~\eqref{eq:k-step-diff} or \eqref{Equ:PLDE}
although with functional or parametric dependences on more variables.
Some algorithms and implementations exist~\cite{Weinzierl:2002hv,Weinzierl:2004bn,Moch:2005uc,Huber:2007dx}
and it will be interesting to pursue
this line of research further within the framework of \pisiSE-fields. 

\medskip

\noindent
{\bf Acknowledgements:}~
This research draws upon (partly still unpublished) results obtained by S.M.
together with A.~Mitov, A.~Vogt and J.~Vermaseren and S.M.
would like to thank them for very pleasant collaborations.
The work of S.M. has been supported by the Helmholtz Gemeinschaft under
contract VH-NG-105 and in part by the Deutsche Forschungsgemeinschaft in
Sonderforschungs\-be\-reich/Transregio~9. 
C.S. was supported by the SFB grant F1305 of the Austrian Science Foundation FWF.

\end{document}